\documentclass[]{jfm}

\usepackage{graphicx}
\usepackage{newtxtext}
\usepackage{newtxmath}
\usepackage{natbib}
\usepackage{hyperref}
\usepackage{xfrac}
\hypersetup{
    colorlinks = true,
    urlcolor   = blue,
    citecolor  = black,
}

\newcommand{\RomanNumeralCaps}[1]
\linenumbers

\title{Absolute and ``upstream'' convective instabilities in plane Couette-Poiseuille flow}

\author{Kirthy K. Srinivas\aff{1}
 \and Sourabh S. Diwan\aff{1}
 \corresp{\email{sdiwan@iisc.ac.in}}}

\affiliation{\aff{1}Department of Aerospace Engineering, Indian Institute of Science, Bengaluru 560012, India}

\begin{document}
\maketitle

\begin{abstract}
Here we report some interesting new features of the spatio-temporal instability of the incompressible plane Couette-Poiseuille flow (CPF). First of all, this flow represents the first instance of a ``non-inflectional'' absolute instability, within constant-viscosity formulation, which is triggered when one of the plates moves opposite to the bulk motion. More strikingly, with further increase in the negative plate motion, the absolute instability ($\textrm{AI}$) transitions to an ``upstream'' convective instability ($\textrm{CI}^-$), wherein an unstable wave packet moves opposite to the direction of the bulk flow. Thus, the CPF exhibits a unique $\textrm{CI}^+ \to \textrm{AI} \to \textrm{CI}^-$ transition, for a given Reynolds number ($Re$), where $\textrm{CI}^+$ denotes the commonly-observed case of a ``downstream'' convective instability. This type of transition has not been reported for other known examples of absolutely unstable flows. We compute the leading and trailing edge velocities for an amplifying wave packet and find that, for the plane Poiseuille flow, both these velocities approach zero as $Re \to \infty$. As a result, at high $Re$, even the slightest of negative plate motions is sufficient to trigger $\textrm{AI}$ and subsequently $\textrm{CI}^-$, as observed for the CPF. The wave-packet dispersion first increases with $Re$, followed by a decrease, which points to a peculiar ``dual'' role of viscosity in sustaining $\textrm{AI}$ in the CPF, namely, viscosity promotes sustenance of $\textrm{AI}$ at moderate Reynolds numbers but suppresses it at low and high Reynolds numbers. These results can be well understood within the Ginzburg-Landau framework, and therefore can be expected to have a wider applicability.
\end{abstract}



\section{Introduction}
\label{sec:intro}

One of the important developments in the modern theory of hydrodynamic stability has been the classification of instabilities into convective and absolute instability \citep{Huerre1990,Chomaz2005}. The convectively unstable flows behave as noise amplifiers, whereas the absolutely unstable flows exhibit an intrinsic oscillator behaviour; the latter can also lead to a ``global'' instability affecting the entire flow domain \citep{Chomaz2005}. 

The prototypical examples of absolutely unstable flows include a mixing layer, a near-wake and a heated jet \citep{Huerre1990}; there have also been several other types of flow, such as separated flows \citep{Alam_Sandham2000}, the rotating-disk boundary layer \citep{Lingwood1995}, the Batchelor vortex \citep{Olendraru2002} etc., in which absolute instability has been observed. It is worth noting that these and other known cases of absolutely unstable flows contain an inflection point in the base velocity profiles and are therefore governed primarily by the inviscid instability mechanism \citep{Huerre1990,Drazin2004}. On the other hand, the non-inflectional flows such as the Blasius boundary layer, the channel flow and the plane Couette flow, which are governed by the viscous instability mechanism \citep{Drazin2004}, have been found to be convectively unstable \citep{Huerre1990}. There exist more complex situations involving absolute instability, such as a channel flow  with viscosity stratification \citep{Rama2014}, but the velocity profile for this case is found to be inflectional. To the best of our knowledge, the onset of absolute instability in a \textit{non-inflectional} flow has not been reported so far. In this work, we show that the plane Couette-Poiseuille flow, within the incompressible constant-viscosity framework, provides the first instance of such a ``viscous non-inflectional'' absolute instability. Interestingly, this flow also exhibits an ``upstream'' convective instability.


The plane Couette-Poiseuille flow (CPF) is obtained by applying favourable pressure gradient to a fluid between two plane parallel plates (Poiseuille component of the flow) and setting one of the plates in motion (Couette component of the flow), and is given by
\begin{equation} \label{base_flow} 
U(y)=\frac{\lambda}{2}(1+y)+(1-y^2) 
\quad
\mathrm{with}
\quad
\begin{aligned} &y = -1, &&U=0\\ &y = +1, &&U=\lambda\end{aligned} 
\end{equation}
where $y$ is the wall normal coordinate, $U$ is the streamwise velocity and $\lambda=\sfrac{U_c}{U_p}$ is ratio of the upper plate speed, $U_c$ and the Poiseuille component of centreline velocity, $U_p$; see figure \ref{fig:abs_rate}(a). Here $y$ is non-dimensionalized by the half-width of the channel, $b$.

The plane Poiseuille flow (PPF), obtained by setting $\lambda=0$ in  (\ref{base_flow}), has been studied extensively as a prototype of linear viscous instability \citep{Drazin2004, Nishioka1975}, as it is a strictly parallel flow that is an exact solution of the Navier-Stokes equations (as is the CPF). \cite{Deissler1987} carried out a spatio-temporal analysis of the PPF (which is the only reported study on the topic) and showed that this flow is convectively unstable at all Reynolds numbers. Interestingly, he reported that the trailing edge velocity of the wave packet decreases with increasing Reynolds number, but the implications of this finding was not explored further \citep{Deissler1987}. This observation, which sets the stage for the present investigation, is found to have far-reaching consequences for the stability of the CPF.


The linear and nonlinear stability characteristics of the CPF have been investigated in the past, both theoretically/numerically \citep{Potter1966,Cowley1985,Kumar2019, Kirthy2021} as well as experimentally \citep{Klotz2017}. An important result from the linear stability analysis \citep{Potter1966} is that the motion of the plate (either in the positive or negative direction) makes the CPF more stable as compared to PPF and for $|\lambda| > 0.7$, there is a complete stabilization of the flow. A physical explanation for this observation was given by \cite{Kirthy2021} in terms of emergence of a region of negative production with an increase in $|\lambda|$; they also carried out a detailed characterization of eigenvalues for $\lambda > 0$ and $\lambda < 0$. Despite these studies, the effect of plate motion on the spatio-temporal behaviour of the CPF has not been reported so far. Here we show that the negative plate speed (i.e. opposite to the Poiseuille component of the flow), although having a slight stabilizing effect on the spatial growth rates, has a profound influence on the convective/absolute character of the CPF - even the slightest reverse flow induced by the moving plate is sufficient to trigger absolute instability, provided the Reynolds number $Re$ is large enough. Here $Re = U_p b/\nu$, $\nu$ being the kinematic viscosity. As the negative plate speed increases, the absolute instability is realized at increasingly lower values of $Re$. Another striking feature we observe is that the absolute instability in the CPF is sustained only for a certain range of $\lambda (<0)$ and $Re$. For $\lambda$ exceeding this range (for a given $Re$), the flow reverts back to being convectively unstable; the direction of the wave-packet propagation, however, is opposite to the bulk fluid motion, i.e., in the upstream direction.

\section{Spatio-temporal stability analysis of Couette-Poiseuille flow}
\label{sec:problem}

The viscous spatio-temporal problem for the evolution of infinitesimal disturbances on a parallel base flow is governed by the Orr-Sommerfeld (O-S) equation \citep{Drazin2004}. For the present base flow (\ref{base_flow}) the O-S equation is solved using the Chebyshev pseudo-spectral method. The disturbances are represented as Fourier modes - $v(x,y,t)=\hat{v}(y)e^{i(\alpha x - \omega t)}$, where $x$ is the streamwise coordinate, $t$ is time, $v$ is the perturbed wall normal velocity, $\hat{v}$ is the complex eigenfunction, and $\alpha$ and $\omega$, respectively, are the complex (streamwise) wavenumber and frequency. 
%
%
Solving the discrete form of the O-S
equation \citep{Kirthy2021} results in a dispersion relation of the form $\omega = \omega(\alpha, Re)$, from which the convective/absolute nature of instability is determined using the Briggs-Bers criterion \citep{Briggs1964,Huerre1985}. This necessitates locating the saddle point of the dispersion relation, $\omega_s$, which is done numerically using the procedure suggested by \cite{Deissler1987}. Our algorithm involves starting with an initial guess $(\alpha_{o},\omega_{o})$ for the saddle point and fitting a complex quadratic to the function $\omega(\alpha)$ using the points $[\omega(\alpha_{o}),\omega(\alpha_{o}+\Delta\alpha_{o}),\omega(\alpha_{o}-\Delta\alpha_{o})]$. By enforcing $\left(\frac{d\omega}{d\alpha}\right)=0$, we obtain a new value for $\alpha_{o}$ which serves as the next guess for the saddle point. This iteration is carried out till the location of the saddle point is calculated to a desired degree of accuracy. For the present work, the eigenvalue computations have been validated against existing results for   the PPF \citep{Schmid2012} and the CPF \citep{Potter1966}; see \cite{Kirthy2021}. The results from the saddle-point computation have been validated against the data of \cite{Deissler1987}; see supplementary figure S1.

\subsection {Absolute and ``upstream'' convective instabilities}

\begin{figure}
  \centering
  \includegraphics[scale=0.65]{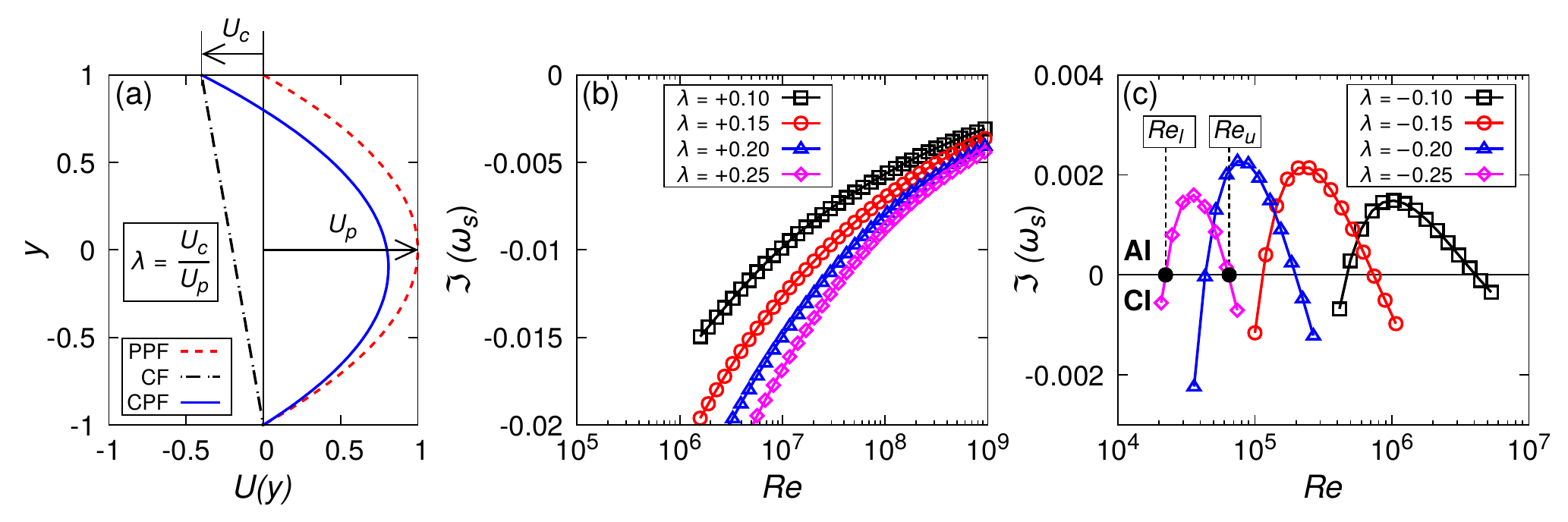}
  \caption{(a) The base velocity profile for the CPF. CF indicates the Couette flow component. (b) The absolute growth rate $\Im(\omega_s)$ as a function $Re$ for $\lambda > 0$, and (c) $\lambda < 0$.}
  \label{fig:abs_rate}
\end{figure}

\begin{figure}
  \centering
  \includegraphics[scale=0.75]{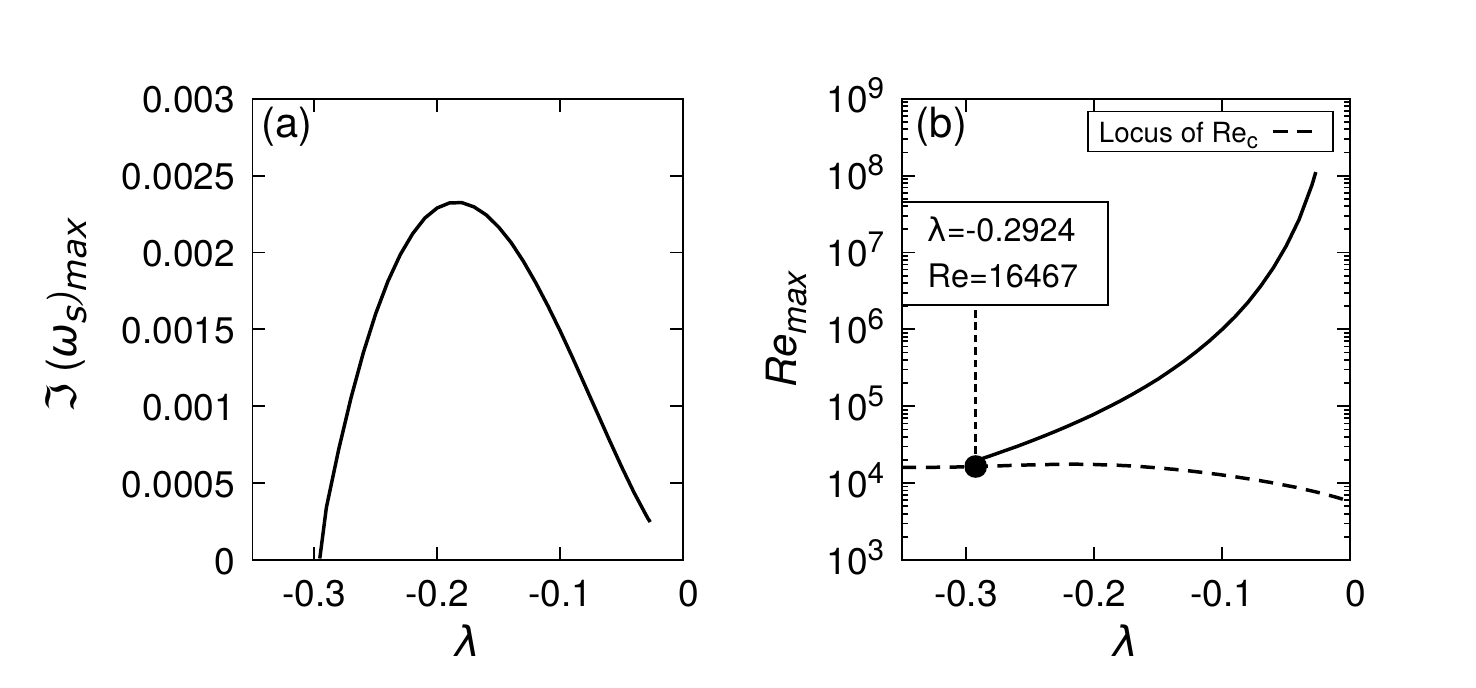}
  \caption{(a) Variation of the maximum absolute growth rate, $\Im(\omega_s)_{max}$, (for a given $\lambda$) as a function of $\lambda$ for the CPF. (b) Variation of the Reynolds number corresponding to the maximum absolute growth rate, $Re_{max}$, as a function of $\lambda$.}
  \label{fig:maxwi}
\end{figure}
\begin{figure}
  \centering
  \includegraphics[scale=0.8]{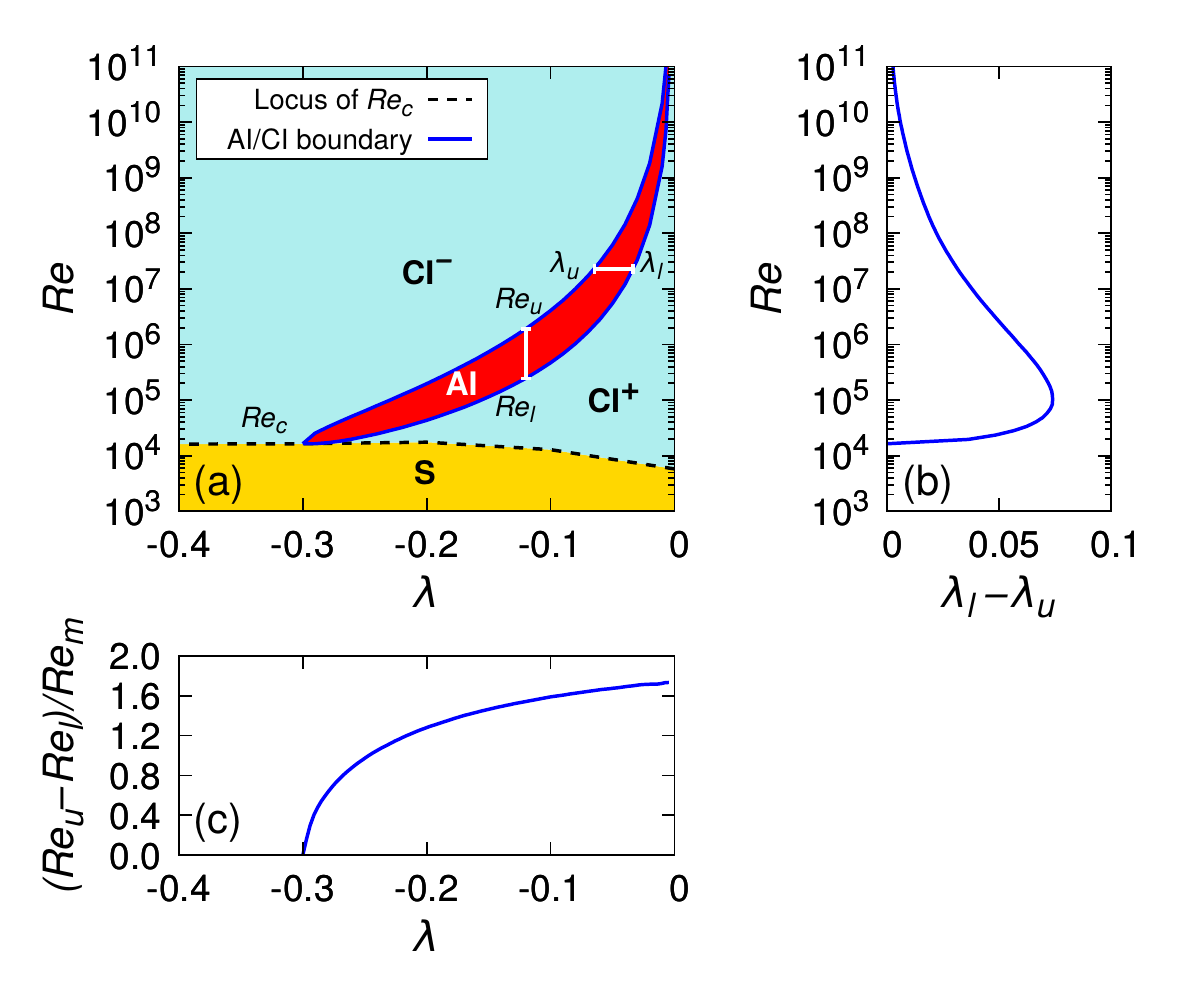}
  \caption{(a) Parametric representation ($Re$, $\lambda$) of the AI/CI instability regions for the CPF. Here $S$ indicates linear stability. (b) The range of $\lambda$, ($\lambda_l-\lambda_u$), supporting AI, as a function of $Re$. (c) The range of $Re$, $(Re_u-Re_l)/Re_m$, supporting AI as a function of $\lambda$; here $Re_m = (Re_u + Re_l)/2$. Note that $Re_u$ and $Re_l$ are not relevant for $\lambda = 0$, which exhibits $\textrm{CI}^+$ for all $Re$ \citep{Deissler1987}.}
  \label{fig:param_space}
\end{figure}

Figures \ref{fig:abs_rate}(b) and \ref{fig:abs_rate}(c) plot the imaginary part of the saddle point, $\Im(\omega_s)$ (which is the temporal growth rate associated with $\omega_s$), as a function of $Re$, respectively, for $\lambda >0$ and $\lambda <0$. 
We find that for $\lambda\in(0,0.7)$ and $Re>Re_c$ ($Re_c$ being the critical Reynolds number), the CPF is convectively unstable, i.e., $\Im(\omega_s)<0$ (figure \ref{fig:abs_rate}b), which is consistent with Deissler's result  for $\lambda=0$ \citep{Deissler1987}. On the other hand, for $\lambda<0$ (and $>-0.2924$; see figure \ref{fig:maxwi}) and for a certain range of $Re$, the flow is found to be absolutely unstable, i.e., $\Im(\omega_s)>0$, as seen from figure \ref{fig:abs_rate}(c). Another interesting observation from figure \ref{fig:abs_rate}(c) is that, for a given $\lambda$, the absolute instability (AI) turns back into convective instability (CI) beyond a particular $Re$. Thus, the CPF is seen to undergo $\textrm{CI}\rightarrow \textrm{AI} \rightarrow \textrm{CI}$ transitions with increase in $Re$; the first transition happens at $Re_l$ and the second at $Re_u$; figure \ref{fig:abs_rate}(c). As $\lambda$ decreases from $-0.1$ to $-0.25$, both $Re_l$ and $Re_u$ go on decreasing (figure \ref{fig:abs_rate}c). For a given $\lambda<0$, the absolute growth rate exhibits a maximum, denoted as $\Im(\omega_s)_{max}$, at the Reynolds number denoted as $Re_{max}$. Figure \ref{fig:maxwi}(a) shows that as $\lambda$ becomes more negative, the (maximum) absolute growth rate first increases and then decreases, reaching zero at $\lambda \approx -0.3$ ($\lambda = -0.2924$ to be precise). For a more negative $\lambda$, $\Im(\omega_s)_{max}<0$, implying transition to a convective instability \citep{Huerre1985}. Thus absolute instability is realized for the CPF only for $-0.2924<\lambda<0$. Interestingly, at $\lambda = -0.2924$, $Re_{max} = Re_c$ and as $\lambda \to 0$, $Re_{max} \to \infty$  (figure \ref{fig:maxwi}b). More comments on this behaviour will be made in relation to figure \ref{fig:param_space}.

\begin{figure}
  \centering
  \includegraphics[scale=0.85]{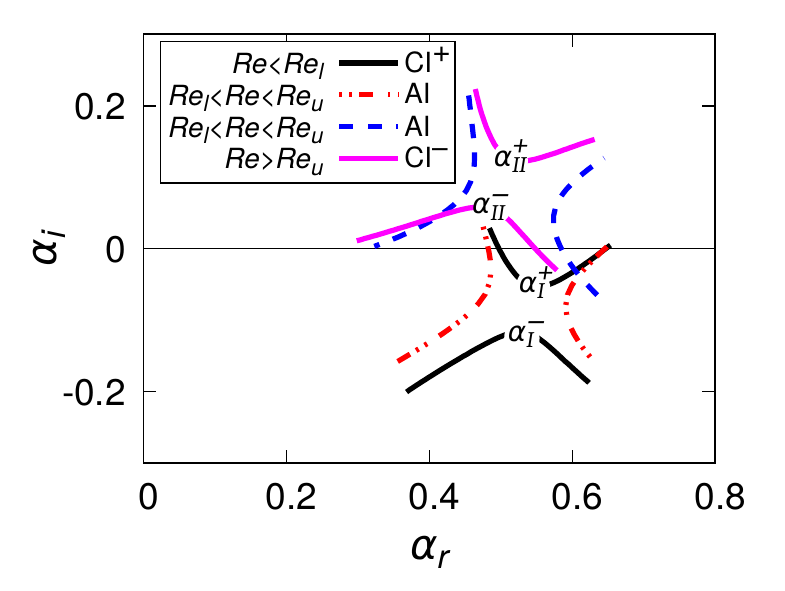}
  \caption{Spatial branches for $\lambda=-0.25$ exhibiting the $\textrm{CI}^{+}\rightarrow \textrm{AI} \rightarrow \textrm{CI}^{-}$ transitions. The four Reynolds numbers used are $2.2 \times 10^4$ ($Re < Re_l$), $2.3 \times 10^4$, $5 \times 10^4$ ($Re_l < Re < Re_u$) and $7 \times 10^4$ ($Re > Re_u$). Note that the branches corresponding to AI do not represent physically meaningful spatial branches as they do not satisfy causality \citep{Huerre1990}; they are included here to track the movement of the saddle point from the lower to upper $\alpha$ planes.}
  \label{fig:kp_km}
\end{figure}

To explore these characteristics further, we plot the spatial branches ($\omega$ real) for $\lambda = -0.25$ in figure \ref{fig:kp_km} for four different  values of $Re$. Here $\alpha^{+}$ represents the downstream travelling branch and $\alpha^{-}$ the upstream branch. The identity of upstream and downstream travelling branches has been confirmed by taking a contour in the complex $\omega$ plane above all the unstable temporal modes. This results in a separation of spatial branches into the lower and upper half $\alpha$ planes \citep{Huerre1985}. For $x>0$ (i.e., the direction of bulk motion), the contour is closed in the upper half $\alpha$ plane and, for $x<0$, in the lower half of $\alpha$ plane (supplementary figure S2). For $Re<Re_l$, the amplifying spatial branch is $\alpha_{I}^{+}$ (figure \ref{fig:kp_km}) which advects in the direction of the bulk flow. This is the well-understood case of ``downstream'' convective instability ($\textrm{CI}^+$), exhibited by a variety of other flows \citep{Huerre1998}. For $Re>Re_u$, however, the spatial branches move into the upper half of the $\alpha$-plane and for this case, $\alpha_{II}^{-}$ becomes the amplifying branch (figure \ref{fig:kp_km}). Thus, for $Re>Re_u$, the unstable wave packet travels upstream, i.e., opposite to the bulk motion, exhibiting an ``upstream'' convective instability ($\textrm{CI}^-$). These features seen for $\lambda=-0.25$ are also observed for the other values of $\lambda$ included in figure \ref{fig:abs_rate}(c). This is shown in supplementary figure S3, in which the movement of the saddle point $(\alpha_s, \omega_s)$ is tracked in the complex $\alpha$ and $\omega$ planes for $\lambda = \pm0.1, \pm0.15, \pm0.2, \textrm{and} \pm0.25$. These results imply that the transitions between $\textrm{CI}$ and $\textrm{AI}$ seen in figure \ref{fig:abs_rate}(c) are of the form $\textrm{CI}^{+}\rightarrow \textrm{AI} \rightarrow \textrm{CI}^{-}$, and these are observed for the the entire range of $\lambda$ for which the CPF is absolutely unstable, i.e., $\lambda \in (-0.2924,0)$. This is evident from the parametric plot presented in figure \ref{fig:param_space}(a).

As seen earlier (figure \ref{fig:abs_rate}c), for a given $\lambda$, $Re_l$ and $Re_u$ represent the $\textrm{CI} -\textrm{AI}$ transition points and are marked in figure \ref{fig:param_space}(a). For $\lambda = -0.2914$, below which $\textrm{AI}$ is not present, $Re_l$ and $Re_u$ coincide with each other and are equal $Re_c$ (figure \ref{fig:maxwi}b). Note that the $\textrm{CI}^{+}\rightarrow \textrm{AI} \rightarrow \textrm{CI}^{-}$ transitions are also observed at a given $Re$ as $\lambda$ becomes more negative (figure \ref{fig:abs_rate}c), with $\lambda_l$ and $\lambda_u$ representing the transition points, as shown in figure \ref{fig:param_space}(a). Thus, for the CPF, the 
$\textrm{AI}$ can be sustained only for a certain range in $\lambda$ and $Re$, namely, $\lambda\in(-0.2924,0)$ and $Re\in(Re_l,Re_u)$, resulting in a ``pocket'' of absolute instability; the ranges in $\lambda$ and $Re$ for which $\textrm{AI}$ is realized are plotted in figures \ref{fig:param_space}(b) and (c) respectively. For a given $\lambda$, the pocket of $\textrm{AI}$ divides the convectively unstable region into two parts - $\textrm{CI}^{+}$ region for $Re\in(Re_c,Re_l)$, and $\textrm{CI}^{-}$ region for $Re\in(Re_u,\infty)$ (figure \ref{fig:param_space}a). A pocket of AI surrounded by a region of CI was also observed for the instability of the Batchelor vortex \citep{Olendraru2002}; however the direction of propagation of CI for this flow was always in the direction of the bulk (axial) flow. The ``upstream'' convective instability, therefore, seems to be a unique feature of the CPF. Interestingly, this upstream convective instability is observed for any $\lambda<0$ (and $> -0.7$), provided $Re>Re_u$; for $\lambda\in(-0.7,-0.2924)$, there exists one continuous $\textrm{CI}^{-}$ region for $Re>Re_c$ (figure \ref{fig:param_space}a). Thus, for $\lambda\in(-0.7,-0.2924)$, the only modal instability possible for the CPF is $\textrm{CI}^{-}$, with the wave packet moving opposite to the bulk motion. (Note that the average velocity for the CPF is $\left(\frac{\lambda}{2}+\frac{2}{3}\right)$ (\ref{base_flow}), which is greater than zero for the entire range of unstable $\lambda \in (-0.7,0)$.)

Another interesting observation from figure \ref{fig:param_space}(a) is that even for \textit{arbitrarily small} negative values of $\lambda$, the CPF exhibits regions of AI and  $\textrm{CI}^{-}$, although at very high values of $Re$. Note that the limit of $Re_l$ and $Re_u$ as $\lambda \to 0$ does not exist, since for $\lambda = 0$, the flow (i.e., the PPF) exhibits only $\textrm{CI}^{+}$ \citep{Deissler1987}. As $\lambda$ becomes more negative for the CPF, $Re_l$ and $Re_u$ drop significantly and approach $Re_{c}$ as $\lambda\rightarrow -0.2924$ (figure \ref{fig:param_space}a). Therefore, it should be possible to realize AI and $\textrm{CI}^{-}$ in the CPF in physical situations at moderate plate speeds, provided the background disturbance levels are sufficiently low to enable modal growth. No such instance has been reported in the literature to our knowledge.

\subsection{Wave-packet propagation and dispersion}

The surprisingly rich behaviour exhibited by the CPF, with regard to the AI-CI transitions, warrants a further investigation. It is well known that, the primary instability mechanism in the CPF, which has a non-inflectional profile, is viscous in character \citep{Drazin2004}. However, the role viscosity plays in promoting and sustaining absolute instability for this flow has not received attention in the literature. To understand this aspect better, we make a Galilean transformation of the observer frame of reference to $x=x_0-Vt$ \citep{Deissler1987}, where $x_0$ represents the laboratory reference frame, and $V$ is the observer speed; see section 2.3 for a discussion on an ``objective'' frame of reference for the CPF. We then proceed to determine the range of $V$ ($=x/t$) for which the wave packet shows temporal growth. The two rays where this growth rate is zero correspond to leading and trailing edges of the amplifying wave packet; see also \cite{Huerre1998}. 

\begin{figure}
  \centering
  \includegraphics[scale=0.75]{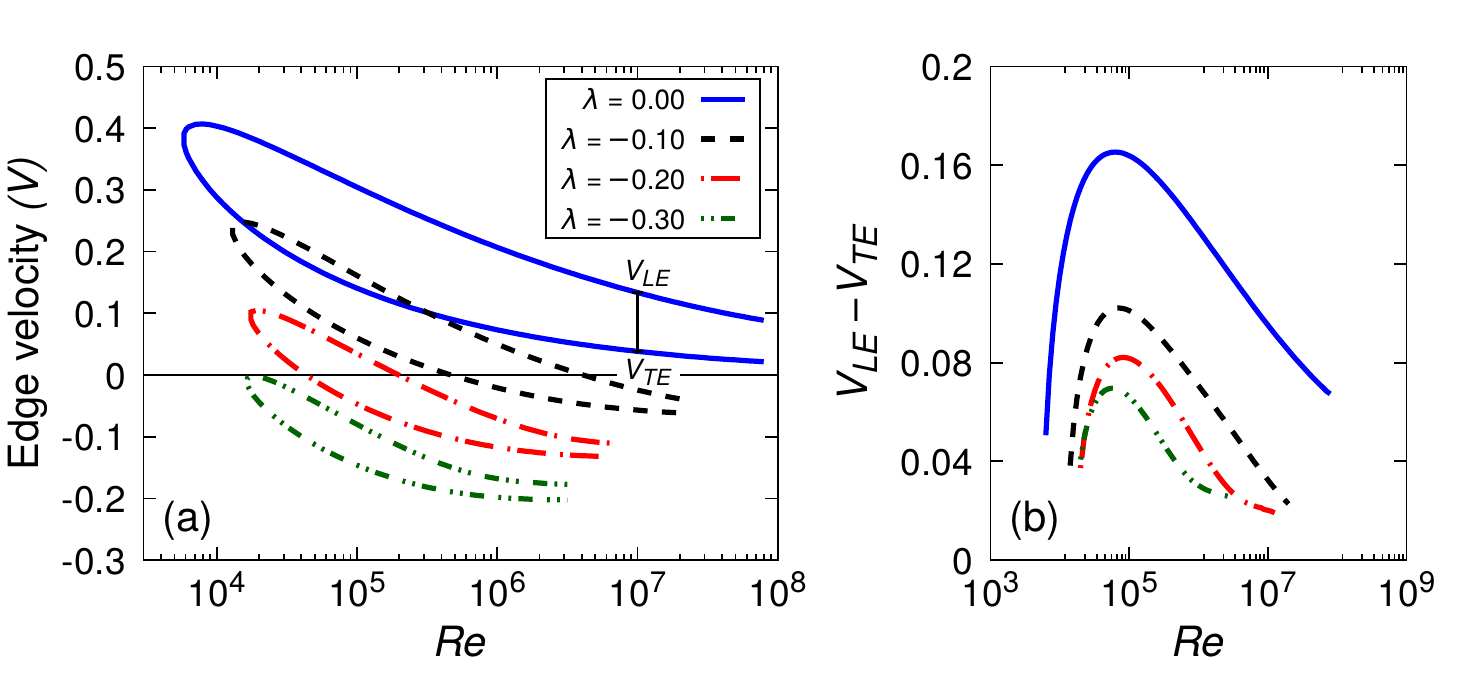}
  \caption{(a) Leading ($V_{LE}$) and trailing edge ($V_{TE}$) velocities of the disturbance wave packet as a function of $Re$. (b) Wave-packet dispersion, $V_{LE}-V_{TE}$, as a function of $Re$.}
  \label{fig:edge_velocities}
\end{figure}

The variation with $Re$ of leading and trailing edge velocities ($V_{LE}$ and $V_{TE}$ respectively) for the CPF is plotted in figure \ref{fig:edge_velocities}(a). We see that $V_{TE}$ decreases monotonically with $Re$ for all $\lambda (\leq 0)$, implying that viscosity has a stabilizing effect on the onset of AI, i.e., a decrease in $Re$ makes it harder for the flow to reach the critical condition for absolute instability. A similar observation has been reported earlier in the context of a wake \citep{Monkewitz1988} and a separated boundary layer \citep{Rodriguez2019}. Interestingly, $V_{TE}\rightarrow 0$ as $Re\rightarrow\infty$ for $\lambda=0$, implying that the PPF is on the verge of becoming absolutely unstable for asymptotically large $Re$ (although it is never realized) and therefore, at these values of $Re$, a slight negative plate speed can make the flow absolutely unstable, as seen in figure \ref{fig:param_space}(a). Furthermore, $V_{LE}$  also decreases monotonically with $Re$. (This was observed by \cite{Deissler1987} for PPF at lower values of $Re$ but he did not calculate $V_{LE}$ for large $Re$ as done presently.) As a result, a further small increase in the negative plate speed can make $V_{LE}$ negative for sufficiently large $Re$. This would make the wave packet switch the direction of propagation and move opposite to the bulk motion. As $Re$ decreases, higher values of negative $\lambda$ are necessary to bring $V_{TE}$ and $V_{LE}$ down to zero and make them negative (figure \ref{fig:edge_velocities}a). Thus the main reason for the switch from AI back to CI (or to $\textrm{CI}^-$) is the significant decrease in $V_{LE}$ with $Re$, which facilitates an upstream propagation of the wave packet for a certain range of negative plate speeds. The peculiar behaviour of $V_{TE}$ and $V_{LE}$ seen in figure \ref{fig:edge_velocities}(a) is what is responsible for the $\textrm{CI}^{+}\rightarrow \textrm{AI} \rightarrow \textrm{CI}^{-}$ transitions observed for the CPF for $-0.2924<\lambda<0$.

The difference between $V_{LE}$ and $V_{TE}$, which is a measure of the degree of dispersion of the wave packet is shown in figure \ref{fig:edge_velocities}(b). For a given $\lambda$, the dispersion first increases with $Re$, reaches a maximum and thereafter decreases.  As a result, for Reynolds numbers corresponding to large dispersion, the $\textrm{AI}$ can be expected to be present for a larger range in $\lambda$, i.e., more negative values of $\lambda$ will be needed to transform $\textrm{AI}$ into $\textrm{CI}^-$. This is evident from figure \ref{fig:param_space}(b), which shows that the range $\lambda_l-\lambda_u$ for which $\textrm{AI}$ exists first increases with $Re$ and then decreases. These results are useful in understanding the role played by viscosity in \textit{sustaining} absolute instability in the CPF. It appears that viscosity plays a ``dual'' role in this regard  - for moderately large $Re$ (corresponding to large wave-packet dispersion; figure \ref{fig:edge_velocities}b), viscosity sustains $\textrm{AI}$ for a larger range of negative $\lambda$ (or ``promotes'' its sustenance; figure \ref{fig:param_space}b), whereas for low and high $Re$, $\textrm{AI}$ is realized over a narrower range of $\lambda$ (implying that viscosity ``suppresses'' the sustenance of $\textrm{AI}$). This behaviour is qualitatively similar to the dual role of viscosity in making the flow unstable in the first place, i.e., the viscous mechanism promotes disturbance amplification only for intermediate values of $Re (> Re_c)$, and the instability gets progressively weaker as $Re \to \infty$ \citep{Drazin2004}. Note that the shapes of the edge velocity curves in figure \ref{fig:edge_velocities}(a) are strikingly similar to those of the neutral curves for the primary instability of CPF \citep{Potter1966, Kirthy2021}. Incidentally, for a fixed $Re$, the wave packet dispersion decreases monotonically as $\lambda$ becomes more negative (figure \ref{fig:edge_velocities}b). This is reflected in a monotonic decrease in the range of $Re$ (i.e., $Re_u-Re_l$) for the realization of $\textrm{AI}$, as $\lambda$ is decreased from zero; see figure \ref{fig:param_space}(c).

\subsection{Choice of an ``objective'' frame of reference for $\textrm{AI}/\textrm{CI}$ distinction}

At this point, it is pertinent to ask whether there exists an ``objective''  laboratory frame of reference for distinguishing between $\textrm{CI}$ and $\textrm{AI}$ for the CPF, as it is a strictly parallel flow. In other words, would the distinction between $\textrm{CI}^+$, $\textrm{AI}$ and $\textrm{CI}^-$ disappear, if the observer moved with a conveniently-chosen frame of reference? As discussed in \cite{Huerre1998}, for a weakly non-parallel flow or for a flow which is continuously forced at a certain location, the laboratory frame of reference is well defined as the streamwise invariance is broken. On the other hand, strictly parallel flows such as the PPF and CPF are invariant with respect to translation in $x$, but the introduction of viscosity requires the no-slip condition to be satisfied, and therefore the laboratory frame of reference for these flows is unambiguously defined. This may be interpreted as the PPF and CPF not being \textit{Galilean invariant} \citep{Huerre1998}, in the sense that any change in the observer frame of reference necessitates the plate velocities to be appropriately changed to satisfy the no-slip condition in the new frame.

To illustrate this point further, we carry out the following exercise. We consider $\lambda<0$, with the upper plate moving in the negative $x$ direction (supplementary figure S4a), which exhibits $\textrm{CI}^-$ for a range of $Re$  (figures \ref{fig:abs_rate}c and \ref{fig:kp_km}). Suppose the observer moves with a reference frame fixed to the upper plate. This makes the lower plate move in the positive $x$ direction, resulting in $\lambda>0$ in the moving frame of reference, with $|\lambda|$ remaining constant (supplementary figures S4b and S4c). As a result of this change in reference frame, the wave-packet edge velocities are shifted by an amount $+|\lambda|$ and the $\textrm{CI}^-$ appears to transform into $\textrm{CI}^+$; see supplementary figure S4(d) for a specific case of $\lambda=-0.25$. (A similar shift in the edge velocities in a moving frame is also seen for the PPF; supplementary figure S5). However, it is important to note that the character of the base flow has changed from $\lambda= -0.25$ in the stationary frame to $\lambda=+0.25$ in the moving frame, and we do not expect $\textrm{CI}^-$ to be present for the latter case since $\lambda>0$ (figure \ref{fig:abs_rate}b). Furthermore, there are important differences in the stability characteristics of the CPF between $\lambda<0$ and $\lambda>0$, with regard to the behaviour of the saddle point (supplementary figure S3) as well as that of the temporal eigenvalues \citep{Kirthy2021}. In summary, an ``objective'' (or the laboratory) reference frame for the CPF for $\lambda<0$ ($\lambda>0$), is any frame in which one of the plates is stationary and the other moves in the opposite (same) direction to the bulk flow. We impose the condition of "zero group velocity" (for determining the saddle point) in this frame of reference for making the $\textrm{AI} / \textrm{CI}$ distinction. If $\lambda <0$, $\textrm{CI}^-$ (as well as $\textrm{AI}$ for $\lambda \in (-0.2924,0)$) is manifested, whereas, if $\lambda >0$, only $\textrm{CI}^+$ is manifested.

\section{Comparison with Ginzburg-Landau framework}
\label{sec:ginzburg}

The $\textrm{CI}^+\rightarrow \textrm{AI} \rightarrow \textrm{CI}^-$ transitions for the CPF can be well understood in terms of the Ginzburg-Landau (G-L) framework. Towards this, we use a slightly modified form of the G-L equation given by
\begin{equation} \label{GL_equation_2}
\frac{\partial\Psi}{\partial t}+\widetilde{U}\frac{\partial\Psi}{\partial x} = \mu \Psi + \kappa \frac{\partial^2\Psi}{\partial x^2},
\end{equation}
where $\Psi (x,t)$ is the disturbance amplitude, $\widetilde{U}$ is velocity, and $\kappa$ is a new parameter introduced to mimic kinematic viscosity in the O-S equation; $\kappa =1$ results in the classical G-L equation \citep{Huerre1990}. Note that $\mu$ in (\ref{GL_equation_2}) does not represent dynamic viscosity but is a measure of disturbance growth. Introducing $\Psi=\hat{\Psi}e^{i(\alpha x - \omega t)}$ into (\ref{GL_equation_2}) and applying the Briggs-Bers criterion gives us the condition for AI as $\mu > \widetilde{U}^2/(4 \:\kappa)$.

\begin{figure}
  \centering
  \includegraphics[scale=0.7]{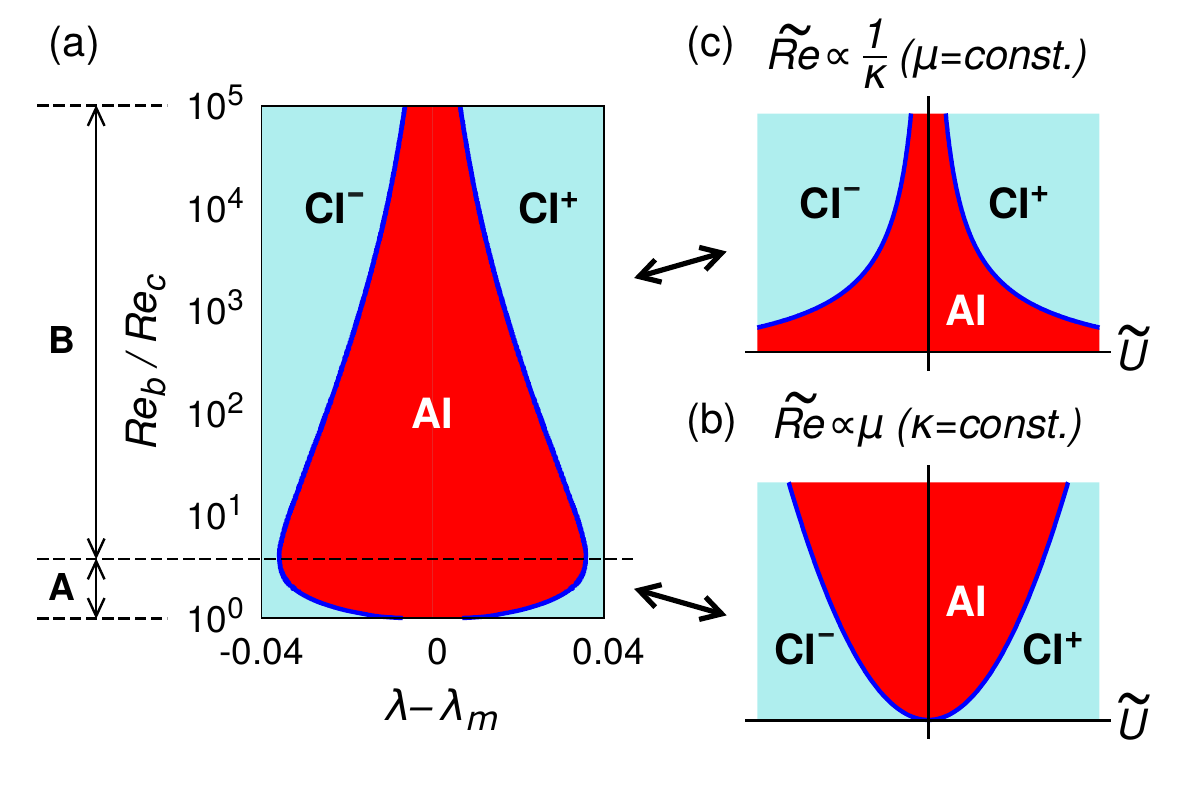}
	\caption{(a) The boundary dividing AI and CI in figure \ref{fig:param_space}(a) represented as $Re_b/Re_c$ as a function of $\lambda-\lambda_m$, where $Re_b$ denotes the Reynolds number corresponding to the dividing boundary and can be either $Re_l$ or $Re_u$, and $\lambda_m=(\lambda_l+\lambda_u)/2$. (b) and (c) The dividing boundary between AI and CI in the $\widetilde{Re}-\widetilde{U}$ space for the G-L equation (\ref{GL_equation_2}), with $\kappa$ and $\mu$ held constant respectively.}
  \label{fig:schematic}
\end{figure}

In (\ref{GL_equation_2}), $\mu$ has the dimensions of velocity gradient and $\kappa$ that of kinematic viscosity. We find it convenient to choose a velocity scale as $\mu L$ and define a Reynolds number given by
\begin{equation} \label{Re_GL}
\widetilde{Re}=\frac{\mu L^2}{\kappa},
\end{equation}
where $L$ is the length scale used to non-dimensionalize $x$ in (\ref{GL_equation_2}), assumed fixed for the present purposes. Note that $\widetilde{U}$ could have also been chosen as a velocity scale for defining $\widetilde{Re}$ but we prefer to use it as a free parameter. We consider $\widetilde{Re}$ and $\widetilde{U}$ to be respectively equivalent to $Re$ and $\lambda$ in the CPF analysis. For enabling comparison, the AI/CI boundary for the CPF on the $Re-\lambda$ plane in figure \ref{fig:param_space}(a) is re-plotted in figure \ref{fig:schematic}(a) in a different representation. 

Next we construct the AI/CI boundary in the $\widetilde{Re}-\widetilde{U}$ space for the modified G-L equation, which we do in two different ways. First, we keep $\kappa$ constant so that $\widetilde{Re}\propto \mu$ (\ref{Re_GL}) and the dividing AI/CI boundary scales as $\widetilde{Re}\propto \widetilde{U}^2$ (figure \ref{fig:schematic}b). Secondly, $\mu$ is held constant so that $\widetilde{Re}\propto (1/\kappa)$ and the dividing boundary in this case is given by $\widetilde{Re} \propto (1/\widetilde{U}^2)$ (figure \ref{fig:schematic}c). It is evident that the first case corresponds to region ``A'' of the dividing boundary for the CPF (figure \ref{fig:schematic}a) where viscosity suppresses the sustenance of $\textrm{AI}$, whereas the second case corresponds to region ``B'' where viscosity promotes the sustenance of $\textrm{AI}$ (by enhancing dispersion as $Re$ decreases; figure \ref{fig:edge_velocities}b). Furthermore, the CI region for the G-L equation is divided into $\textrm{CI}^+$($\widetilde{U}>0$) and $\textrm{CI}^-$($\widetilde{U}<0$). This is evident from the sign of the imaginary part of the absolute wavenumber for the G-L equation ($\alpha_s=-\frac{i\widetilde{U}}{2\kappa}$) \citep{Huerre1998}. When $\widetilde{U}$ changes from positive to negative, $\alpha_s$ moves from lower half to upper half in the complex $\alpha$ plane, implying that the wave packet amplifies in the negative $x$ direction; see supplementary figure S6. This analysis shows that the G-L equation reproduces the CI-AI transitions for the CPF (i.e, $\textrm{CI}^+ \to \textrm{AI} \to \textrm{CI}^-$) more closely than it does for other common examples, such as a mixing layer or a separated boundary layer, which exhibit only the $\textrm{CI}^+ \to \textrm{AI}$ transition; the transition from AI to $\textrm{CI}^{-}$ for these flows has not been reported to our knowledge \citep[see, e.g.,][]{Huerre1985}. This provides a further support to the utility of the G-L equation as a model problem for studying the spatio-temporal instability of fluid flows.

\section{Conclusion}\label{sec:filetypes}
We have reported a spatio-temporal stability analysis of the Couette-Poiseuille flow using the Orr-Sommerfeld equation to reveal some remarkable and previously unknown features with regard to the $\textrm{AI} / \textrm{CI}$ behaviour of the CPF, as summarized below.
\begin{enumerate}
    \item The CPF is shown to exhibit absolute instability for a range of negative plate speeds, $\lambda \in (-0.2924,0)$. Even the slightest of negative plate motions is sufficient to trigger $\textrm{AI}$, provided $Re$ is large enough.
    
    \item For a given $Re$, the CPF undergoes  $\textrm{CI}^+ \to \textrm{AI} \to \textrm{CI}^-$ transitions, as $\lambda$ is made progressively more negative. The presence of an ``upstream'' convective instability ($\textrm{CI}^-$), i.e., an unstable wave packet moving opposite to the bulk motion, seems to be a unique feature of the CPF not reported previously.
    
    \item For $\lambda \in (-0.7, -0.2924)$, the only instability possible for the CPF is the ``upstream'' convective instability. Neither $\textrm{CI}^+$ nor $\textrm{AI}$ is realized for these values of $\lambda$.
    
    \item For the PPF, the leading and trailing edge velocities of the unstable wave packet decrease monotonically with $Re$,  approaching zero as $Re \to \infty$. It is this feature that enables realization of $\textrm{AI}$ and $\textrm{CI}^-$ for the CPF, even for small negative values of $\lambda$.
    
    \item For a given $\lambda$, the wave-packet dispersion first increases with $Re$, following which it decreases. This reveals a peculiar ``dual'' role of viscosity in sustaining absolute instability in the CPF; viscosity promotes sustenance of $\textrm{AI}$ at moderate Reynolds numbers but suppresses it for lower and higher values of $Re$.
    
    \item The $\textrm{CI}^+ \to \textrm{AI} \to \textrm{CI}^-$ transitions observed for the CPF are consistent with the Ginzberg-Landau (G-L) framework, which also admits similar transitions. In fact, the CPF exemplifies the G-L model better than other known examples of linearly unstable flows.
\end{enumerate}

 We expect some of these results to be applicable to other non-inflectional flows, governed by viscous mechanism, that might become absolutely unstable. This can potentially open up new directions of research in understanding the role of viscosity in triggering and sustaining absolute instability in shear flows.


\backsection[Acknowledgements]{The authors thank Dr. Abhijit Mitra from Israel Institute of Technology, Technion, for lending his stability code. Thanks are due to Prof O. N. Ramesh from  IISc, Bengaluru, for useful discussions.}

\backsection[Funding]{SSD acknowledges financial support from IISc, Bengaluru as a start-up grant (No. 1205010620).}

\backsection[Declaration of interests]{The authors report no conflict of interest.}

\bibliographystyle{jfm}
\bibliography{stab_lit}

\end{document}